\documentclass[aps,prl,twocolumn,superscriptaddress,longbibliography]{revtex4-2}
\usepackage{graphicx}
\usepackage{amssymb,amsmath}
\usepackage{bm}
\usepackage{dcolumn}
\usepackage{float}
\usepackage[OT1]{fontenc} 
\usepackage{url}
\usepackage{mathrsfs}
\usepackage{slashed,comment}
\usepackage{color}
\usepackage{verbatim}
\usepackage{enumitem}
\usepackage{soul,physics}
\usepackage[driverfallback=dvipdfm]{hyperref}
\hypersetup{pdfpagemode=FullScreen,colorlinks=true,breaklinks,urlcolor=blue,linkcolor=blue,citecolor=blue}

\usepackage{subfigure}
\usepackage{amssymb}
\usepackage{bm}
\usepackage{graphicx}
\usepackage{amsmath,amssymb}

\begin{document}

%%%%%% title
\title{Designing Shadow Tomography Protocols by Natural Language Processing}

%%%%%% authors

\author{Yadong Wu}
\affiliation{College of Physics, Sichuan University, Chengdu 610065, China}

\author{Pengfei Zhang}
\affiliation{Department of Physics \& State Key Laboratory of Surface Physics, Fudan University, Shanghai 200438, China}
\affiliation{Hefei National Laboratory, Hefei 230088, China}

\author{Ce Wang}
\affiliation{School of Physics Science and Engineering, Tongji University, Shanghai, 200092, China}

\author{Juan Yao}
\affiliation{International Quantum Academy, Shenzhen 518048, Guangdong, China}

\author{Yi-Zhuang You}
\email{yzyou@physics.ucsd.edu}
\affiliation{Department of Physics, University of California at San Diego, La Jolla, CA 92093, USA}

%\author{Hui Zhai}
%\email{huizhai.physics@gmail.com}
%\affiliation{Institute for Advanced Study, Tsinghua University, Beijing 100084, China}
%\affiliation{Hefei National Laboratory, Hefei 230088, China}

% make title
\date{\today}

\begin{abstract}
Quantum circuits form a foundational framework in quantum science, enabling the description, analysis, and implementation of quantum computations. However, designing efficient circuits, typically constructed from single- and two-qubit gates, remains a major challenge for specific computational tasks. In this work, we introduce a novel artificial intelligence–driven protocol for quantum circuit design, benchmarked using shadow tomography for efficient quantum state readout. Inspired by techniques from natural language processing (NLP), our approach first selects a compact gate dictionary by optimizing the entangling power of two-qubit gates. We identify the {\sf iSWAP} gate as a key element that significantly enhances sample efficiency, resulting in a minimal gate set of \{{\sf I}, {\sf SWAP}, {\sf iSWAP}\}. Building on this, we implement a recurrent neural network trained via reinforcement learning to generate high-performing quantum circuits. The trained model demonstrates strong generalization ability, discovering efficient circuit architectures with low sample complexity beyond the training set. Our NLP-inspired framework offers broad potential for quantum computation, including extracting properties of logical qubits in quantum error correction.
\end{abstract}

\maketitle

\textbf{Introduction.--} 
Quantum computing presents a paradigm shift in computation, offering significant advantages over classical systems by addressing problems currently intractable for conventional approaches \cite{PhysRevLett.127.180501,Arute:2019aa,doi:10.1126/science.abn7293,Daley:2022ab}. Typically, quantum computing is formulated using quantum circuits, which are composed of interconnected quantum gates. These circuits exhibit varying levels of expressivity and find broad application across diverse fields such as quantum physics, cryptography, and computer science. The efficiency of a quantum circuit is fundamentally determined by two key aspects: its structure, which defines the arrangement of quantum gates, and the parameters chosen for each gate. Different circuit architectures possess distinct expressive capabilities, while tuning of gate parameters is essential for optimizing performance on specific computational tasks \cite{Saeedi:2011aa,PhysRevResearch.3.L032057,PhysRevLett.132.010602,Du:2022aa,PhysRevLett.101.060401,PhysRevResearch.3.L032049,childs_et_al:LIPIcs.TQC.2019.3}. 

Once a quantum circuit's structure and parameters are established, it's expected to provide accurate predictions for its intended applications. However, the design of optimal quantum circuits is a non-trivial challenge. Even for quantum circuits with highly expressive structures, determining the optimal parameters of quantum gates can be computationally demanding. Gradient-based optimization methods often face challenges such as the barren plateau phenomenon in deep quantum circuits \cite{Larocca:2025aa,McClean:2018aa}. This underscores the critical need for more efficient and robust methods for quantum circuit generating. 

Notably, quantum circuits can be conceptualized as temporal sequences of quantum gates. This inherent sequential characteristic makes them amenable to techniques from natural language processing (NLP), a rapidly evolving field well-suited to modeling and generating sequences \cite{Chowdhary2020,doi:10.1126/science.aaa8685}. Consequently, leveraging NLP frameworks for the design and optimization of quantum circuits offers a promising and innovative approach. Although recent studies have begun to explore the development of quantum circuit generators for certain basic quantum algorithms, such efforts are often limited to highly restricted scenarios \cite{10.1145/3519939.3523433,Pointing_2024,Benedetti:2019aa,fosel2021quantumcircuitoptimizationdeep}. The broader challenge of generating quantum circuits for complex and general tasks remains largely unresolved, highlighting the need for a more systematic integration of artificial intelligence methodologies into quantum circuit design.

\begin{figure}
    \begin{center}  
    \includegraphics[width=0.98\linewidth]{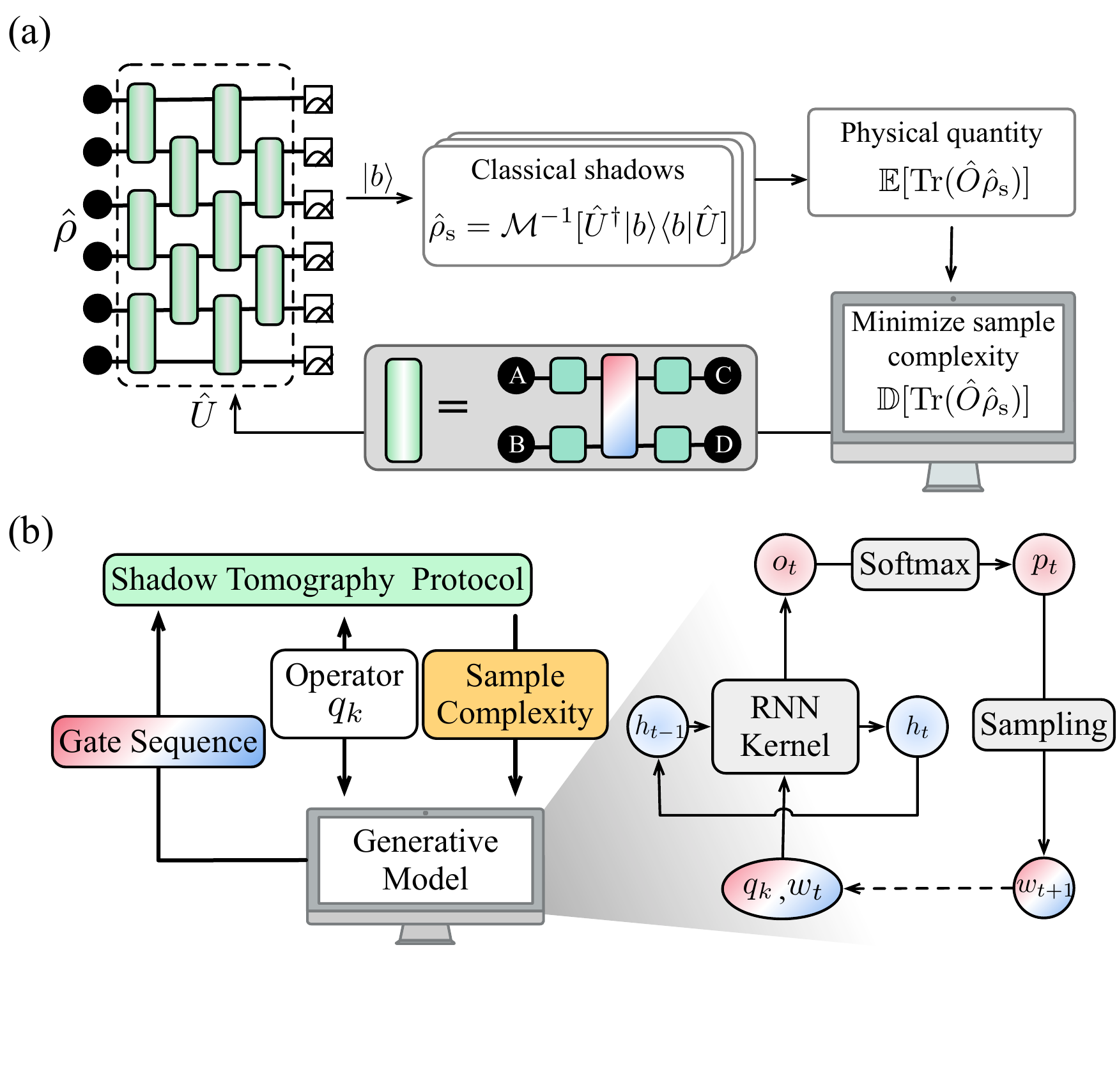}
    \caption{ (a) Optimizing shadow tomography efficiency with locally-scrambled invariant circuits. An unknown quantum state ($\hat{\rho}$) undergoes evolution via a random circuit ($\hat{U}$) and is then measured in the computational basis ($|b\rangle$). Classical shadows reconstructed from these measurements provide unbiased predictions of Pauli expectations. Optimizing the random circuit directly improves the sample efficiency for extracting these quantities. (b) Training a gate-sequence generative Model using a NLP framework. A RNN within the model generates quantum circuits from input operators. The shadow tomography protocol then evaluates these generated circuits, providing feedback in the form of sample complexity. This feedback is used to train the generative model, with the goal of minimizing the required sample complexity.
    \label{fig:shadow}}
    \end{center}
\end{figure}

In this work, we demonstrate the advantages of applying an NLP framework for quantum circuit generation by focusing on shadow tomography as a concrete example \cite{Huang:2020aa}. This protocol utilizes randomized measurements and has emerged as a highly efficient technique for simultaneously predicting multiple physical properties of an unknown quantum state—an essential capability for characterizing large-scale quantum systems. As illustrated in FIG.~\ref{fig:shadow}(a), the protocol involves applying a unitary $\hat{U}$ randomly sampled from an ensemble $\mathcal{E}_U$ to an unknown quantum state $\hat{\rho}$. The system is then measured in the computational basis $|b\rangle$, and a reconstruction channel $\mathcal{M}^{-1}$ is applied to obtain classical shadows: $\hat{\rho}_{\rm s} = \mathcal{M}^{-1}[\hat{U}^\dagger |b\rangle\langle b| \hat{U}]$. These classical shadows enable unbiased estimation of observable expectations, $\mathrm{Tr}(\hat{O}\hat{\rho}) = \mathbb{E}[\mathrm{Tr}(\hat{O}\hat{\rho}_{\rm s})]$, where the variance, $\Vert\hat{O}\Vert^2_{\mathcal{E}_U}=\mathbb{D}[{\rm Tr}(\hat{O}\hat{\rho}_{\rm s})]$, quantifies the protocol's sample complexity. This complexity is critically dependent on the chosen unitary ensemble. For a locally-scrambled invariant unitary ensemble \cite{PhysRevB.102.134203,PhysRevB.101.224202,PhysRevResearch.5.023027}, the sample complexity is related to the operator size distribution of the evolved operators $\hat{O}_U=\hat{U}\hat{O}\hat{U}^\dagger$ as follows \cite{PhysRevLett.130.230403,qi2019measuringoperatorsizegrowth,Bu:2024aa}:
\begin{equation}
\Vert\hat{O}\Vert^2_{\mathcal{E}_{U}}=w(\hat{O}_U)_{\mathcal{E}_{U}}^{-1}, \quad w(\hat{O}_U)_{\mathcal{E}_{U}}=\sum_m\frac{\pi(m)_{\mathcal{E}_{U}}}{3^{m}},
\end{equation}
where $\pi(m)$ denotes the size distribution of $\hat{O}_U$ and $w(\hat{O}_U)_{\mathcal{E}_U}$ is known as the Pauli weight.

While recent efforts have proposed novel random unitary ensembles to reduce the sample complexity of shadow tomography by incorporating prior human knowledge, these protocols primarily focus on local operators supported on contiguous regions \cite{PhysRevLett.133.020602,Akhtar2023scalableflexible,PRXQuantum.2.030348,doi:10.1126/science.abk3333,PhysRevB.109.094209,PRXQuantum.5.010350,zhang2024holographicclassicalshadowtomography,PhysRevResearch.4.013054,Zhou2024efficientclassical,akhtar2025dualunitaryshadowtomography,Hu:2025aa,wu2024contractiveunitaryclassicalshadow,PhysRevResearch.6.013029,Ippoliti2024classicalshadows}. Efficient methods for observables defined on arbitrary non-local subsystems, however, remain elusive. To address this limitation, we implement an NLP framework to train a gate-sequence generative model capable of addressing more general measurement scenarios. Considering the NLP model predicts two-qubit gates from a discrete set, named {\it dictionary}, our methodology begins by reducing the dictionary dimensionality of two-qubit gates, through which we identify the {\sf iSWAP} gate as crucial for significantly improving sample efficiency. Subsequently, we employ a recurrent neural network (RNN) as the sequence generator. To train this generative model with sample complexity and gate complexity as the evaluation metric, we utilize a policy gradient RL algorithm. The converged model successfully generates efficient random unitary circuits for extracting physical observables with arbitrary supports.

{\bf Dictionary Construction.--}
General two-qubit gates belong to the SU(4) group, a continuous Lie group with 15 generators. Directly using this infinite set as an NLP dictionary for training generative models is computationally infeasible. Therefore, an important initial step for gate-sequence generating tasks is to refine the gate-dictionary. We focus on random quantum circuits composed of locally scrambled two-qubit gates, which have been widely adopted in shadow tomography protocols. Each two-qubit gate can be decomposed as $\hat{U}_{i,i+1}^t=u_iu_{i+1}\Lambda_{i,i+1}^tv_iv_{i+1}$, as depicted in FIG.~\ref{fig:shadow}(a). Here, $\{u,v\}$ adhere to the circular unitary ensemble (CUE), $t=1,\ldots,L$ denotes the circuit layers, and $i=1,\ldots,N-1$ indexes the qubits. This decomposition implies that the local-basis dependence of individual two-qubit gates is not critical, thereby reducing the effective degrees of freedom of the unitary $\hat{\Lambda}$ \cite{PhysRevA.63.062309}.

To enable efficient generative modeling, we further extract a discrete subset of gates that are most impactful for shadow tomography. To identify this subset, we express the unitary operator $\hat{\Lambda}$ in its Choi representation, which transforms it into a pure state over systems A, B, C, and D, shown in FIG.~\ref{fig:shadow}(a). 
\begin{figure}
    \begin{center}  
    \includegraphics[width=0.98\linewidth]{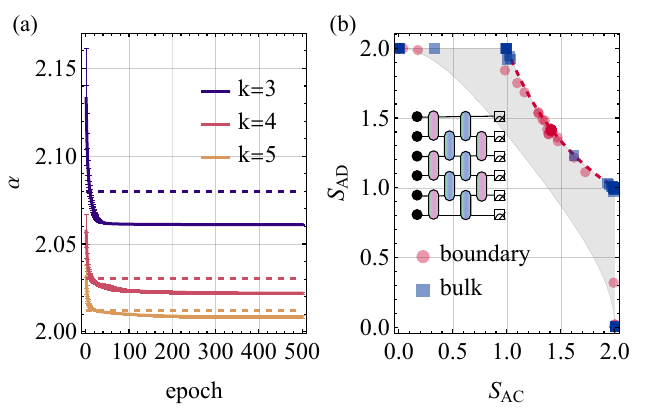}
    \caption{  (a) Optimizing the second-order purities of two-qubit gates when extracting successive operators of size $k$. The curves represent the average scaling parameters $\alpha_k = (\mathbb{D}[{\rm Tr}\hat{O}_k\hat{\rho}_{\rm s}])^{1/k}$, averaged over 10 different initializations, converging to lower values than the random Clifford protocol's scaling, $\alpha_{k,\rm RC} = (2^k+1)^{1/k}$ (dashed lines), indicating improved efficiency. (b) R\'enyi Entropies of optimized two-qubit gates trained in (a). Red circles indicate gates at the circuit boundaries, while blue squares represent gates within the bulk of the circuit (see inset). The shaded region represents the allowed second-order purities for these gates. Notably, bulk gates converge to the four vertices of this shaded region, and boundary gates converge along the red dashed line,  contributing equivalently to shadow tomography.
    \label{fig:EFoptim}}
    \end{center}
\end{figure}
 Our study focuses on successive Pauli operators with sizes $k=3, 4, 5$ whose corresponding circuit layers are $L=4, 6, 8$. We employed Adam's method to optimize the second-order purities $\rm Tr\hat{\rho}_{AC}^2$ and $\rm Tr\hat{\rho}_{AD}^2$ of two-qubit gates, initiating the training process from 10 distinct random initializations to ensure robust convergence and explore the parameter landscape effectively. The training outcomes are visually presented in FIG.~\ref{fig:EFoptim}(a). For successive operators, the scaling parameter $\alpha_k \triangleq (\mathbb{D}[\mathrm{Tr}(\hat{O}_k \hat{\rho}_{\mathrm{s}})])^{1/k}$ converges to a significantly lower value compared to that when unitaries $\hat{U}_{\rm RC}$ are sampled from the conventional random Clifford ensemble on the support of the operator. This optimized ensemble possesses distinct entanglement properties from the Clifford ensemble. 

Following the training, we conducted an analysis of the second Rényi entropies, $S_{\mathrm{AC}} = -\log_2 \mathrm{Tr}(\hat{\rho}_{\mathrm{AC}}^2)$ and $S_{\mathrm{AD}} = -\log_2 \mathrm{Tr}(\hat{\rho}_{\mathrm{AD}}^2)$, for the two-qubit gates. FIG.~\ref{fig:EFoptim}(b) allowed us to identify the gates that contribute most significantly to the shadow tomography protocol.  The gates within the bulk of the random circuits converge to the vertices of the allowed region \cite{PhysRevB.98.014309,kong2024convergenceefficiencyquantumgates} , specifically corresponding to the (locally-scrambled) {\sf I, SWAP, iSWAP}, and {\sf CZ} gates. In contrast, boundary gates converged not only to these four vertices but also to points along one specific boundary, indicated by the red dashed line in FIG.~\ref{fig:EFoptim}(b). We rigorously proved that all boundary gates along this red shaded line are equivalent in sample complexity contribution \cite{supp}. Since the {\sf I} gate and the {\sf SWAP} gate introduce no entanglement, they do not alter the operator size distribution. And the {\sf CZ} gate is equivalent to the composition of a {\sf iSWAP} gate and a {\sf SWAP} gate. Therefore, the {\sf iSWAP} gate plays the dominant role in the operator size contraction, which not only increase the probability of operator $\hat{O}_U$ with small size, but also connect un-successive qubits. This aligns with recent research highlighting the advantages of these gates in achieving efficient information scrambling and realizing random Haar ensembles in 2-designs \cite{kong2024convergenceefficiencyquantumgates}.

\textbf{Reinforce Training with RNN.--}Building on the optimized two-qubit gate dictionary $\mathcal{D} = \{ \textsf{I}, \textsf{SWAP}, \textsf{iSWAP}\}$, our goal is to generate quantum circuits that minimize the sample complexity for arbitrary target Pauli operators. We first encode the support of a Pauli operator $\hat{O}_{q_k}$ as a binary vector  $q_k \in \{0,1\}^N$, where $q_k^{(i)} = 1$ indicates the operator acts non-trivially on the $i_{\rm th}$ qubit . We adopt a two-layer vanilla RNN as the core sequence generator, tailored for modeling quantum gate sequences shown in FIG.~\ref{fig:shadow}(b). At each timestep $t$, the RNN takes two inputs: the support vector $q_k$ and the gate configuration $w_t$. $w_t$ is formed by concatenating $N_g$ one-hot vectors with dimension $|\mathcal{D}|$, where $N_g$ is the number of two-qubit gates per layer and the dimension of $w_t$ is $|\mathcal{D}|\times N_g$. These inputs drive the hidden state update $h_t = f(h_{t-1}, [q_k, w_t])$, where $f$ is the RNN cell.  The output layer $o_t$ subsequently generates a probability distribution $p_t$ over the gate dictionary $\mathcal{D}$ for sampling gates at layer $t+1$. Here we denote all trainable parameters as $\theta$. Note that while our circuit adopts a brick-wall structure, the inclusion of \textsf{SWAP} gates renders it equivalent to an arbitrary architecture given sufficient circuit depth. This process is underpinned by a reinforcement learning (RL) paradigm, where the RNN functions as an agent interacting with a quantum environment \cite{sutton1998reinforcement,ranzato2015sequence}. In this environment, unitary operations evolve the system's state. The state space of the RL agent is defined by the current layer's gate configuration, $w_t$. The agent's actions select unitary evolutions $w_{t+1}$ for the next time step according to a policy $p_t$, subsequently updating its internal state $h_t$. A reward signal is provided only after the agent generates a complete sequence $\mathbf{w}={w_1,\cdots,w_L}$, which represents a quantum circuit. This reward is based on the circuit's sample complexity for a random $q_k$, with a small penalty added for gate complexity. The reward function is defined as:
\begin{align}
r(q_k,\mathbf{w})=r(\Vert\hat{O}{q_k}\Vert^2_{\rm RNN})-r_g(\mathbf{w})
\end{align}
where the first term represents the sample complexity and the second term is the function of the number of {\sf SWAP} gates in $\mathbf{w}$, characterizing the gate complexity. The objective is to maximize the expected reward $\mathbb{E}_{\mathbf{w} \sim p_\theta}[r(q_k,\mathbf{w})]$. Accordingly, we define the loss function as the negative expected reward, which is approximated in practice by averaging over $N_a$ sampled sequences.  
 \begin{align}
 \mathcal{L}_\theta = -\mathbb{E}_{\mathbf{w} \sim p_\theta}[r(q_k,\mathbf{w})] \approx -\frac{1}{N_a}\sum_{a=1}^{N_a} r(q_k,\mathbf{w}^a)
 \end{align}

\begin{figure}
    \begin{center}  
    \includegraphics[width=0.9\linewidth]{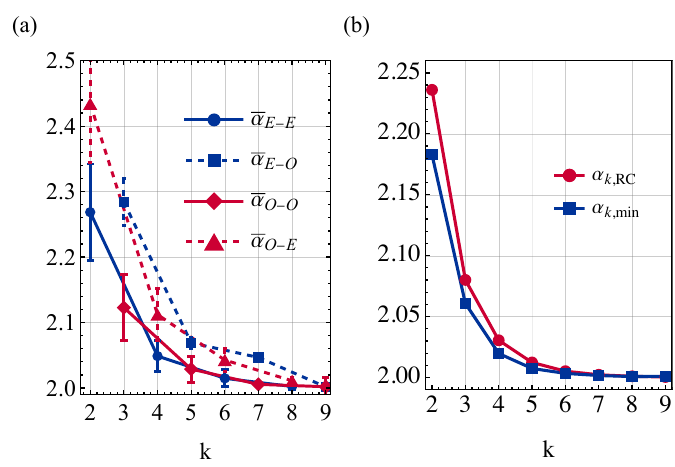}
    \caption{ (a) The average scaling parameters ($\alpha$) for a shadow tomography protocol using $N=9$ qubits. Solid curves illustrate the generative model's effectiveness when trained and tested on the same operator set (E-E for $\hat{O}_{k_e}$, O-O for $\hat{O}_{k_o}$), showcasing its learning capability. The dashed lines reveal the generator's generalization power by comparing performance when training and testing operator sets are different (e.g., E-O trains on $\hat{O}_{k_e}$ and tests on $\hat{O}_{k_o}$). (b) Comparing generator-predicted minimum scaling with random Clifford. The red solid lines depict the scaling parameters $\alpha_{k,\rm RC}$ achieved by the random Clifford protocol applied to the successive size $k$ operator. The blue solid line presents the minimal scaling parameter predicted by the generative model, demonstrating its ability to find more efficient shadow tomography protocols.
        \label{fig:RNNpre}}
    \end{center}
\end{figure}

To expedite convergence and leverage this exploration, we implement an experience replay mechanism \cite{kapturowski2018recurrent}.  The training is hybrid: at the primary phase, the RL explores gate sequences, updates parameters via policy gradient algorithm. At the periodic supervised phase, we update the RNN by supervised learning the experience buffer to reproduce high-performance sequences. Further details regarding the training methodology are provided in the supplementary material \cite{supp}.

\textbf{Results.--}
We investigate a quantum system comprising $N=9$ qubits. The generative model, based on our NLP framework, is trained on Pauli operators with restricted subsystem sizes. Training is conducted independently for two distinct classes of operator supports with even number of supports $k_e \in \{2, 4, 6, 8\}$ or odd number of supports $k_o \in \{3, 5, 7\}$. Given that for a specific support set $k_{e/o}$, there are $n_k=\sum_{k\in k_{e/o}}\left(\begin{matrix}N\\k\end{matrix}\right)$ different configurations. The generator is trained aiming to predict circuits for extracting Pauli operators with any supports $q_k$. Post-training, the generator not only constructs efficient quantum circuits for operators within its training set but also exhibits remarkable generalization capabilities to unseen support sizes. The scaling parameter is defined as $\alpha_{q_k}=(\Vert\hat{O}_{q_k}\Vert^2_{})^{1/k}$. The average and minimal scaling parameter of support size $k$ are written as:
\begin{equation}
\bar{\alpha}_k=\frac{1}{n_k}\sum_{q_k}\alpha_{q_k},~~~~~\alpha_{k,\min}=\min_{q_k}\alpha_{q_k}.
\end{equation} 
FIG.~\ref{fig:RNNpre}(a) explicitly demonstrates the agent's capacity to generate high-quality solutions. For test operators with odd (even) supports evaluated using models trained on operators with even (odd) supports, denoted as E-O (O-E), the scaling parameter still converges to low values. This performance is notably superior to that of the conventional random Clifford shallow circuit protocol, where scaling parameters are in the regime $\alpha_{\rm sc}\in[2,2.28]$ for successive operators \cite{PhysRevLett.130.230403}. Furthermore, FIG.~\ref{fig:RNNpre}(b) shows that for a specific $k$, the generator predicts quantum circuits with a minimal sample complexity in all $q_k$, lower than that to the random Clifford protocol. This numerically validates that our generator can significantly improve sample efficiency beyond the typical random Clifford case and more predictions are shown in the supplementary material \cite{supp}. 

\begin{figure}
    \begin{center}  
    \includegraphics[width=0.85\linewidth]{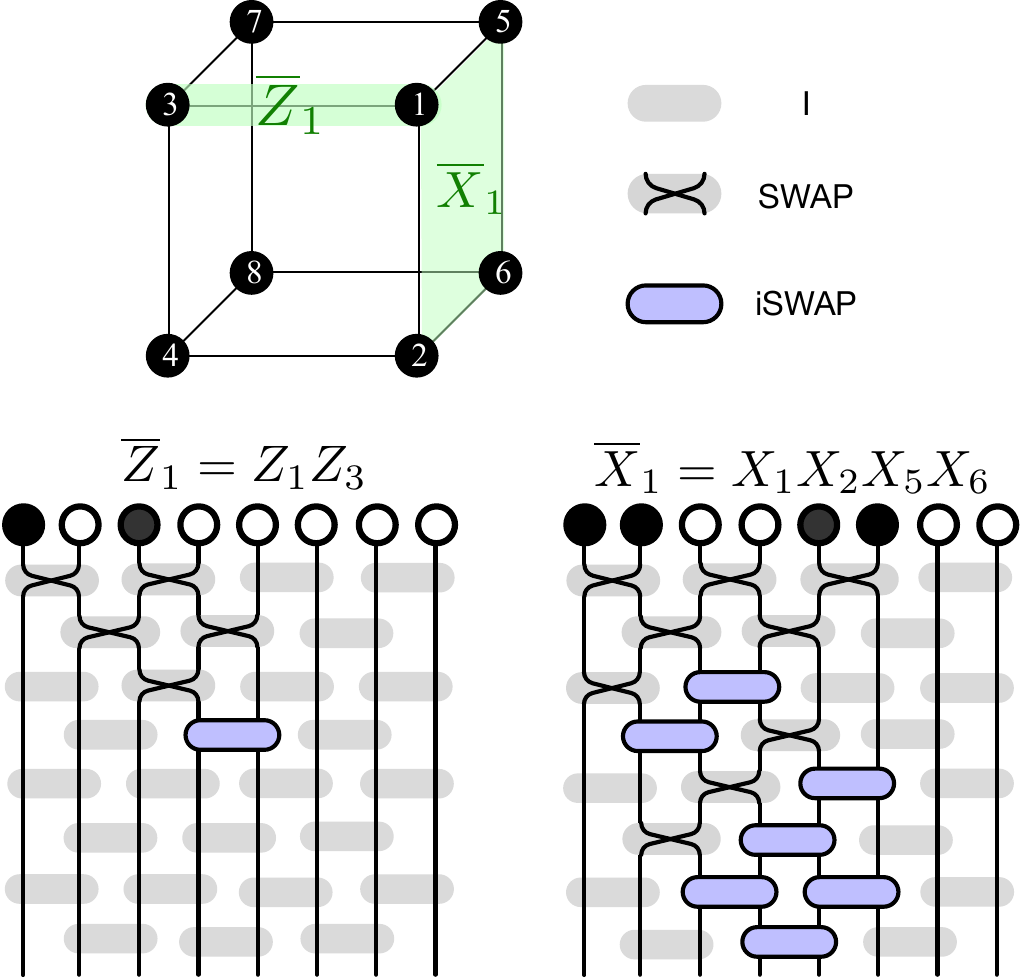}
    \caption{ The generator's predicted circuits for efficiently extracting a non-local operators beyond the training set. The generator trained with $k_o$ while the logical operators of [[8,3,2]] color code are supported at even number of qubits. The predicted circuits have small gate complexity and the scaling parameters obtained from the generator are smaller than  those achieved with the random Clifford protocol i.e. $\alpha_{\overline{Z}_1}=2.183<\alpha_{k=2,\rm RC}=2.236,~ \alpha_{\overline{X}_1}=2.020<\alpha_{k=4,\rm RC}=2.031$.
    \label{fig:OpCir}}
    \end{center}
\end{figure}

This gate-sequence generator predicts resource-efficient quantum circuits for arbitrary operator supports, enabling direct application in the efficient extraction of non-local properties, such as logical properties in quantum error correction (QEC). In quantum error correction \cite{RevModPhys.87.307,lidar2013quantum}, quantum information is encoded in logical qubits using redundant physical qubits to achieve noise tolerance. However, extracting physical quantities necessitates measuring logical operators, which are highly non-local when expressed in terms of physical operators, such as logical operator in quantum low-density parity check code (qLDPC)\cite{PRXQuantum.2.040101}. Here we consider the smallest 3D color code as an example \cite{Kubica_2015}, whose physical qubits lie on vertices of a cube.  As depicted in FIG.\ref{fig:OpCir}, the generator, trained with odd supports, successfully generates circuits for predicting logical operators with even-support more efficiently, comparing with the scaling of random Clifford protocols. This QEC protocol can be directly realized in trapped-ion, atom array and superconducting devices \cite{Bluvstein:2024aa,PhysRevA.109.062438,doi:10.1126/sciadv.ado9024}.

\textbf{Discussions.--}
This work introduces a novel concept for constructing quantum circuits using artificial intelligence. Specifically, we leverage a NLP framework to train a gate-sequence generative model for shadow tomography, a protocol known for efficiently extracting quantum information from unknown states. We first refined the candidate set of two-qubit gates. This dictionary cleaning process revealed that the {\sf iSWAP} gate significantly contributes to improving the sample efficiency of the shadow tomography protocol. After this data cleaning process, our gate dictionary was reduced to a compact set comprising three essential elements: {\sf I, SWAP, iSWAP}. Then, Utilizing RNNs as the sequence generator and optimizing parameters through a policy gradient method within a RL paradigm, our generative model can predict efficient quantum circuits for extracting arbitrary Pauli operators. This work contributes to the growing interplay between machine learning and quantum information science, demonstrating that NLP techniques can be effectively leveraged to interpret and generate quantum gate sequences. 

While this work focuses on random unitary ensembles with a brick-wall structure, the NLP-based learning framework is broadly adaptable to other quantum circuit architectures. By treating the circuit structure as a learnable output rather than a fixed input, the framework can jointly optimize gate sequences and topologies, potentially uncovering novel layouts beyond human intuition. This not only improves the sample efficiency of shadow tomography but also opens new directions for AI-driven quantum circuit design. For example, the generative model can be applied to tasks like quantum state preparation, quantum machine learning and quantum error correction\cite{Benedetti_2019,Biamonte:2017aa,balasubramanian2025localautomaton2dtoric}. In the context of quantum neural networks, it may help discover expressive, trainable ansatzes with fewer parameters. Moreover, the framework’s modular design allows for future integration with experimental constraints such as finite error rates, enhancing its practicality for near-term quantum devices.

\vspace{5pt}
\textit{Acknowledgement.} 
We thank Hui Zhai for his helpful discussions.  This project is supported by the NSFC under grants No.12504309(YW), No.12374477 (PZ), No.12204352(CW), the Shanghai Committee of Science and Technology grant No.25LZ2600800(YW), the Shanghai Rising-Star Program under grant No.24QA2700300 (PZ), the Innovation Program for Quantum Science and Technology 2024ZD0300101 (PZ), the Science, Technology and Innovation Commission of Shenzhen, Municipality KQTD20210811090049034 (JY), and Guangdong Basic and Applied Basic Research Foundation 2022B1515120021 (JY). YZY is supported by a startup fund from UCSD.

\bibliography{refShadowSeq}

\end{document}